\documentclass[%
 reprint,
 amsmath,amssymb,
 aps,
]{revtex4-2}

\usepackage{graphicx}
\usepackage{dcolumn}
\usepackage{bm}
\usepackage{hyperref}

\begin{document}

\preprint{APS/123-QED}

\title{On the extension of double copy procedure to higher derivative double field theory}

\author{Rasim Yılmaz}%
\email{ryilmaz@metu.edu.tr}
\affiliation{%
 Department of Physics, Faculty of Arts and Sciences\\
 Middle East Technical University, 06800, Ankara, Türkiye
}%

\date{\today}

\begin{abstract}
Double field theory (DFT) can be constructed from the color-kinematic double copy of Yang-Mills theory. In a recent work, this construction has been extended to higher-derivative terms, starting from the four-derivative extension of Yang-Mills theory, to obtain a conformally invariant DFT action up to the third order. Here, I attempt to extend this idea by introducing a method, inspired by the background independent formulation of DFT, to obtain third order higher derivative terms directly from the second order higher derivative terms. The third-order terms I obtain do not match those obtained directly from the double-copy map. A clear understanding of this mismatch can give valuable information about the double copy procedure for DFT, its relation to background independence, and the conformal symmetry in double configuration space.

\end{abstract}

\maketitle

\section{Introduction}
Double Field Theory (DFT) is a proposal to incorporate T-duality, a distinctive symmetry of string theory, as a symmetry of a field theory defined on a double configuration space \cite{dft}. The theory introduces dual coordinates to the usual coordinates and doubles the number of dimensions to have a manifest $O(D,D)$ invariance, which can be understood as a more general version of T-duality. In double configuration space, all the objects can be written with $O(D,D)$ indices, and the action can be written in a manifestly $O(D,D)$ invariant way by using the generalized metric and the dilaton field \cite{dftgenmetric}. The underlying geometry of DFT can also be written in an $O(D,D)$ invariant way by using the corresponding connections and covariant derivatives, which are introduced in \cite{RiemDFT} and \cite{invariantgeometry}. Other important works about the underlying geometry of DFT are \cite{framelike,projcomp,stringydiffgeo,doubledgeo}. However, the symmetries in DFT cannot determine the Riemann tensor uniquely, and there remain some undetermined parts \cite{RiemDFT}. Imposing further symmetries can be useful in determining these parts uniquely. For instance, conformal symmetry can be used as a tool. However, we have a limited understanding of conformal symmetry in double configuration space, so it does not seem easy to construct a conformally invariant DFT action and compare it with conformal gravity. At this point, it is better to consider other approaches to give us new insights about DFT.

The Bern-Carrasco-Johansson (BCJ) double copy \cite{bcj}, which relates the scattering amplitudes of gauge and gravity theories, is one of these new approaches that gives new insights into both gauge and gravity theories. The idea is based on KLT relations \cite{stringdc}, which tells that closed string tree-level amplitudes can be written in terms of open-string amplitudes. These double copy relations for scattering amplitudes are also extended to the relations between classical solutions of Yang-Mills theory and gravity (see, for example, \cite{blackholesdc,taubnut,maxsymm}). In an interesting recent work \cite{jaram}, the authors show that a precise double copy prescription for the Yang-Mills action at quadratic and cubic order yields DFT action in which the duality invariant dilaton has been integrated out. The double structure of DFT, which suggests a relation with the double copy idea, inspired other works such as \cite{kerrschilddc,heterokerr,heterolescan}.

The idea that DFT action can be constructed as a double copy of Yang-Mills action has recently been extended to higher-derivative actions. In \cite{lescan}, the authors started with the four-derivative extension of Yang-Mills theory and obtained a four-derivative DFT action up to cubic order by applying the double copy prescription. The motivation of this work was the observation that the double copy of specific higher-derivative Yang-Mills theories corresponds to conformal (super)gravity \cite{johansson}. This construction is extended in the work \cite{lescannew} by including a charged scalar field to the double copy procedure. The authors showed that the action they obtained reduces to the conformal gravity action at quadratic order in the pure gravity limit. Moreover, this action in pure gravity limit at the cubic order reproduces terms in the Weyl gravity action for some particular choices and gauge fixing conditions. Including higher derivative terms, this action should be related to the DFT at order $\alpha'$, whose geometry is first examined in \cite{Hohm:2013jaa}. In \cite{Hohm:2014xsa}, \cite{Marques:2015vua}, higher derivative DFT up to order $\alpha'$ is constructed as a two-parameter family of theories. Different choices of these parameters lead to different higher derivative theories. In \cite{Hohm:2014xsa}-\cite{Marques:2015vua}, the higher derivative DFT with $Z_2$ invariance is called as $DFT+$ and the $Z_2$ odd theory is called as $DFT-$. Moreover, in \cite{Hohm:2014xsa}, it has been shown that $DFT-$ can be written in a manifestly background independent way in terms of a metric formulation, whereas the gauge algebra will be field dependent in a manifestly background independent formulation of $DFT+$.

In this work, I attempt to extend this higher-derivative DFT action obtained in \cite{lescan} to higher orders by using the expansions of the background independent objects. I start with the second-order terms of the higher-derivative DFT action and extend these terms to higher-order terms, using a method similar to the one introduced in \cite{hohmback}. Although it was expected to get at least the third order terms by this construction, my cubic order terms do not match the cubic order terms obtained in \cite{lescan} by using the double copy prescription. 

In section 2, I start by reviewing the construction \cite{jaram} of the DFT action from a double copy prescription on the Yang-Mills action. Then, I review the results of \cite{lescan} in section 3 and derive the higher derivative DFT action from a double copy map. In section 4, I give the background independent formulation of DFT \cite{hohmback}, which will be the starting point for the method employed in the rest of the work. Then, in section 5, I obtain the third order terms of higher derivative DFT action by starting from the second order terms in \cite{lescan} and apply the method in \cite{hohmback}. Next, I show that the cubic terms obtained are inconsistent with the results of \cite{lescan}. Finally, I discuss the possible causes and implications of the results.

\section{DFT action as the double copy of Yang-Mills action}
In this section, I review the construction \cite{jaram} of the DFT action as a double copy map of the Yang-Mills action. The starting point is the action for the Yang-Mills theory in $D$-dimensions
\begin{equation}
    S_{\mathrm{YM}}=-\frac{1}{4} \int d^D x \, \kappa_{a b} \, F^{\mu \nu}{ }^a \, F_{\mu \nu}{ }^b,
\label{yangmills}
\end{equation}
with the Yang-Mills field strength
\begin{equation}
    F_{\mu \nu}{ }^a=\partial_\mu A_\nu{ }^a-\partial_\nu A_\mu{ }^a+g_{\mathrm{YM}} f^a{ }_{b c} A_\mu{ }^b A_\nu{ }^c,
\end{equation}
 where ${{g_{\mathrm{YM}}}}$ is the coupling constant and $f^a{ }_{b c}$ denotes the structure constants of the gauge group.
This action can be shown to be equal to
\begin{equation}
    S_{\mathrm{YM}}^{(2)} = \frac{1}{2} \int d^D x \, \kappa_{a b} A^{\mu a}\left( \square A_\mu{ }^b-\partial_\mu \partial^\nu A_\nu{ }^b\right)
\end{equation}
 up to a total derivative term. Before applying the double copy procedure, it is convenient to transform to momentum space. Then the Yang-Mills action can be written as
\begin{equation}
    S_{\mathrm{YM}}^{(2)}=-\frac{1}{2} \int d^D k \, \kappa_{a b} \, k^2 \, \Pi^{\mu \nu}(k) A_\mu{ }^a(-k) A_\nu{ }^b(k),
\end{equation}
where the projector is defined in the Minkowski space $\eta_{\mu \nu}= (-,+,+,+)$ as
\begin{equation}
    \Pi^{\mu \nu}(k) \equiv \eta^{\mu \nu}-\frac{k^\mu k^\nu}{k^2}.
    \label{projector}
\end{equation}
Note that the factor $k^2$ is scaled out before the projector is defined, which will later become clear in the double copy procedure. The first step in the double copy prescription is to replace the color indices with the dual spacetime indices $(a \longrightarrow \Bar{\mu})$ corresponding to the dual set of spacetime momenta $\Bar{k}^{\Bar{\mu}}$:
\begin{equation}
    A_\mu{ }^a(k) \rightarrow e_{\mu \bar{\mu}}(k, \bar{k}) .
    \label{adc}
\end{equation}
The Cartan-Killing metric should also be replaced, and the double copy prescription for it is given by
\begin{equation}
    \kappa_{a b} \longrightarrow \frac{1}{2} \bar{\Pi}^{\bar{\mu} \bar{\nu}}(\bar{k}),
    \label{strucdc}
\end{equation}
where the dual projector $\bar{\Pi}^{\bar{\mu} \bar{\nu}}(\bar{k})$ is defined in analogy to (\ref{projector}) with all the momenta and indices replaced by their barred versions. Note that the $k^2$ factors are not replaced with anything in this double copy prescription. This is consistent with the usual double copy idea because color factors are replaced with kinematic factors while the propagators remain fixed in the usual double copy prescription for amplitudes.  \\
By using this prescription, the action up to quadratic order is found as
\begin{equation}
    S_{\mathrm{DC}}^{(2)}=-\frac{1}{4} \int_{k, \bar{k}} \, k^2 \, \Pi^{\mu \nu}(k) \, \bar{\Pi}^{\bar{\mu}\bar{\nu}}(\bar{k}) \, e_{\mu \bar{\mu}}(-k,-\bar{k}) \, e_{\nu \bar{\nu}}(k, \bar{k}),
    \label{momentdc}
\end{equation} 
with the notation $\int_{k, \bar{k}} \equiv \int d^Dk d^D{\Bar{k}}$. This action is obviously symmetric in the usual momenta $k$ and dual momenta $\Bar{k}$ except for the $k^2$ factor. However, this does not break the symmetry because the construction of DFT includes the level-matching condition, which  states that 
\begin{equation}
    k^2 = {\Bar{k}}^2.
\end{equation} 
It is necessary to return to the position space to relate this action to the usual DFT action in quadratic order. By using the definition of the projector given in (\ref{projector}) and the corresponding definition for the dual projector, the action (\ref{momentdc}) can be written as 
\begin{equation}
\begin{aligned}
    S_{\mathrm{DC}}^{(2)}=-\frac{1}{4} \int_{k, \bar{k}} &\left(k^2 e^{\mu \bar{\nu}} e_{\mu \bar{\nu}}-k^\mu k^\rho e_{\mu \bar{\nu}} e_\rho^{\bar{\nu}}-\bar{k}^{\bar{\nu}} \bar{k}^{\bar{\sigma}} e_{\mu \bar{\nu}} e^\mu{ }_{\bar{\sigma}} \right. \\
    &\left. +\frac{1}{k^2} k^\mu k^\rho \bar{k}^{\bar{\nu}} \bar{k}^{\bar{\sigma}} e_{\mu \bar{\nu}} e_{\rho \bar{\sigma}}\right) .
\end{aligned}
\label{momentumdclast}
\end{equation} 
Since the last piece yields a non-local term when transformed to the position space, it is necessary to introduce an auxiliary scalar (dilaton) field $\phi(k,\Bar{k})$ as
\begin{equation}
    \phi=\frac{1}{k^2} k^\mu \bar{k}^{\bar{\nu}} e_{\mu \bar{\nu}}. 
\end{equation} 
By using this auxiliary field, (\ref{momentumdclast}) becomes 
\begin{equation}
\begin{aligned}
    S_{\mathrm{DC}}^{(2)}=-\frac{1}{4} \int_{k, \bar{k}} ( & k^2 e^{\mu \bar{\nu}} e_{\mu \bar{\nu}}-k^\mu k^\rho e_{\mu \bar{\nu}} e_\rho^{\bar{\nu}}-\bar{k}^{\bar{\nu}} \bar{k}^{\bar{\sigma}} e_{\mu \bar{\nu}} e^\mu{} _{{\bar{\sigma}}} \\
    &  -k^2 \phi^2+2 \phi k^\mu \bar{k}^{\bar{\nu}} e_{\mu \bar{\nu}}).
\end{aligned}
\end{equation} 
Now, Fourier transforming to the position space, the action becomes 
\begin{equation}
\begin{aligned}
    S_{\mathrm{DC}}^{(2)}=\frac{1}{4} \int d^D x & d^D \bar{x}\left(e^{\mu \bar{\nu}} \square e_{\mu \bar{\nu}}+\partial^\mu e_{\mu \bar{\nu}} \partial^\rho e_\rho{ }^{\bar{\nu}} \right.\\
    &\left.+\bar{\partial}^{\bar{\nu}} e_{\mu \bar{\nu}} \bar{\partial}^{\bar{\sigma}} e^\mu{ }_{\bar{\sigma}}-\phi \square \phi+2 \phi \partial^\mu \bar{\partial}^{\bar{\nu}} e_{\mu \bar{\nu}}\right),
\end{aligned}
\label{dcquad}
\end{equation} 
which corresponds to the quadratic part of the full DFT action. To extend this construction to the cubic order, consider the cubic part of the Yang-Mills action (\ref{yangmills}) 
\begin{equation}
    S_{\mathrm{YM}}^{(3)}=-g_{\mathrm{YM}} \int d^D x f_{a b c} \, \partial^\mu A^{\nu a} \, A_\mu^b A_\nu^c.
\end{equation} 
By a Fourier-transform to momentum space, this becomes 
\begin{equation}
    S_{\mathrm{YM}}^{(3)}=\frac{i g_{{YM}}}{(2 \pi)^{D / 2}} \int_{k_i} \delta\left(k_1+k_2+k_3\right) f_{a b c} k_1^\mu A_1^{\nu a} A_{2 \mu}{ }^b A_{3 \nu}{ }^c
\end{equation} 
with the notation $A_i \equiv A(k_i)$. This can be written more symmetrically as 
\begin{equation}
\begin{aligned}
    S_{\mathrm{YM}}^{(3)}= & -\frac{i g_{\mathrm{YM}}}{6(2 \pi)^{D / 2}}  \int_{k_1, k_2, k_3}  \delta\left(k_1+k_2+k_3\right) \\
    \times & f_{a b c}\pi^{\mu \nu \rho}\left(k_1, k_2, k_3\right) A_{1 \mu}{ }^a A_{2 \nu}{ }^b A_{3 \rho}{ }^c,
    \label{cubicmom}
\end{aligned}
\end{equation} 
where the projector is defined as 
\begin{equation}
    \pi^{\mu \nu \rho}\left(k_1, k_2, k_3\right) \equiv \eta^{\mu \nu} k_{12}^\rho+\eta^{\nu \rho} k_{23}^\mu+\eta^{\rho \mu} k_{31}^\nu,
    \label{cubicproj}
\end{equation} 
with $k_{i j} \equiv k_i-k_j$. The double copy prescription for the structure constant is given as 
\begin{equation}
    f_{a b c} \rightarrow \frac{i}{4} \bar{\pi}^{\bar{\mu} \bar{\nu} \bar{\rho}}\left(\bar{k}_1, \bar{k}_2, \bar{k}_3\right).
    \label{structuredc}
\end{equation} 
Using (\ref{structuredc}) and (\ref{cubicproj}) in (\ref{cubicmom}), together with $g_{\mathrm{YM}} \rightarrow \frac{1}{2} \kappa$ and setting $\kappa=1$,
\begin{equation}
\begin{aligned}
S_{\mathrm{DC}}^{(3)}= & \frac{1}{8(2 \pi)^{D / 2}} \int  d K_1 d K_2 d K_3 \delta\left(K_1+K_2+K_3\right)\\
&e_{1 \mu \bar{\mu}}\left(-k_2^\mu e_{2 \rho \bar{\rho}} \bar{k}_3^{\bar{\mu}} e_3^{\rho \bar{\rho}}+k_2^\mu e_{2 \nu \bar{\rho}} \bar{k}_3^{\bar{\rho}} e_3^{\nu \bar{\mu}}+k_2^\rho e_2^{\mu \bar{\rho}} \bar{k}_3^{\bar{\mu}} e_{3 \rho \bar{\rho}}\right. \\
& \left.+k_2^\mu \bar{k}_2^{\bar{\mu}} e_{2 \rho \bar{\rho}} e_3^{\rho \bar{\rho}}-k_{2 \rho} e_2^{\mu \bar{\mu}} \bar{k}_{3 \bar{\rho}} e_3^{\rho \bar{\mu}}-k_2^\rho \bar{k}_2^{\bar{\mu}} e_2^{\mu \bar{\rho}} e_{3 \rho \bar{\rho}}\right),
\end{aligned}
\end{equation} 
with the notation $e_{i \mu \bar{\mu}} \equiv e_{\mu \bar{\mu}}\left(K_i\right)$, $K \equiv (k,\bar{k})$, and $d K \equiv d^{2 D} K$. 
Since no part can yield a non-local term after Fourier transformation, it is not necessary to define an auxiliary field and one can directly go to the position space. The action in position space, after integrating by parts, reads 
\begin{equation}
\begin{aligned}
    S_{\mathrm{DC}}^{(3)}=&\frac{1}{8} \int_{x, \bar{x}} \, e_{\mu \bar{\mu}}\left(2 \partial^\mu e_{\rho \bar{\rho}} \bar{\partial}^{\bar{\mu}} e^{\rho \bar{\rho}}-2 \partial^\mu e_{\nu \bar{\rho}} \bar{\partial}^{\bar{\rho}} e^{\nu \bar{\mu}} \right. \\
    & \left. -2 \partial^\rho e^{\mu \bar{\rho}} \bar{\partial}^{\bar{\mu}} e_{\rho \bar{\rho}}+\partial^\rho e_{\rho \bar{\rho}} \bar{\partial}^{\bar{\rho}} e^{\mu \bar{\mu}}+\bar{\partial}_{\bar{\rho}} e^{\mu \bar{\rho}} \partial_\rho e^{\rho \bar{\mu}}\right).
\end{aligned}
\label{dccubic}
\end{equation} 
In \cite{jaram}, this action has been shown to be equal to the cubic DFT action by imposing a gauge-fixing condition and integrating out the dilaton.
\section{Higher derivative DFT Action as double copy of $(DF)^2$ Yang-Mills action}
In this section, I review the higher derivative extension of the DFT action by using a double copy map from the higher derivative Yang-Mills action following \cite{lescan}. A higher derivative extension of the usual Yang-Mills theory is given by the Lagrangian 
\begin{equation}
    \mathcal{L}=\frac{1}{2} \kappa_{a b} \, D_\mu F^{\mu \nu a} \, D_\rho F^\rho{ }_\nu{ }^b,
    \label{higherymlag}
\end{equation} 
where the gauge covariant derivative is defined as 
\begin{equation}
    D_\rho F_{\mu \nu}{ }^a=\partial_\rho F_{\mu \nu}{ }^a+g_{\mathrm{YM}} f^a{ }_{b c} A_\rho{ }^b F_{\mu \nu}{ }^c .
\end{equation} 
The usual double copy prescription for this higher-derivative extension of Yang-Mills theory has been shown to correspond to conformal gravity at the level of amplitudes \cite{johansson}. The question is whether one can find a conformally invariant DFT action by using the double copy of this Lagrangian. The Lagrangian (\ref{higherymlag}) up to quadratic order can be written as
\begin{equation}
    S_{\mathrm{HD}}^{(2)}=\frac{1}{2} \int d^D x \, \kappa_{a b} \, \square A^{\mu a}\left(\square A_\mu{ }^b-\partial_\mu \partial^\nu A_\nu{ }^b\right) .
\end{equation}
To impose the double copy prescription in this action, it is necessary to go to the momentum space by writing this action as
\begin{equation}
    S_{\mathrm{HD}}^{(2)}=-\frac{1}{2} \int_k \kappa_{a b} k^4 \Pi^{\mu \nu}(k) A_\mu{ }^a(-k) A_\nu{ }^b(k).
    \label{highermom}
\end{equation}
Imposing the double copy maps given by (\ref{adc}) and (\ref{strucdc}), the action (\ref{highermom}) becomes
\begin{equation}
    S_{\mathrm{HDC}}^{(2)}=-\frac{1}{4} \int_{k, \bar{k}} k^4 \Pi^{\mu \nu}(k) \bar{\Pi}^{\bar{\mu} \bar{\nu}}(\bar{k}) e_{\mu \bar{\mu}}(-k,-\bar{k}) e_{\nu \bar{\nu}}(k, \bar{k}),
\end{equation} 
which has the same form with (\ref{momentdc}) except for the $k^2$ factor. Expanding the projectors, this action becomes 
\begin{equation}
\begin{aligned}
    S_{\mathrm{HDC}}^{(2)}=-\frac{1}{4} \int_{k, \bar{k}} k^2(&k^2 e^{\mu \bar{\nu}} e_{\mu \bar{\nu}}-k^\mu k^\rho e_{\mu \bar{\nu}} e_\rho^{\bar{\nu}}-\bar{k}^{\bar{\nu}} \bar{k}^{\bar{\sigma}} e_{\mu \bar{\nu}} e^\mu{ }_{\bar{\sigma}} \\
    & +\frac{1}{k^2} k^\mu \bar{k}^{\bar{\nu}} k^\rho \bar{k}^{\bar{\sigma}} e_{\mu \bar{\nu}} e_{\rho \bar{\sigma}}).
    \label{highdcquadmom}
\end{aligned}
\end{equation}
Note that the extra $k^2$ factors will cancel the non-local terms in the last piece, and there will be no need to introduce an auxiliary field while going to the position space. After transforming to the doubled position space, the action takes the form
\begin{equation}
\begin{aligned}
S_{\mathrm{HDC}}^{(2)}=-\frac{1}{4} &\int d^D x d^D \bar{x} \left(\square e^{\mu \bar{\nu}} \square e_{\mu \bar{\nu}}-\square e^{\mu \bar{\nu}} \partial_\mu \partial^\rho e_{\rho \bar{\nu}} \right.\\
& \left. -\square e^{\mu \bar{\nu}} \bar{\partial}_{\bar{\nu}} \bar{\partial}^{\bar{\sigma}} e_{\mu \bar{\sigma}}+\partial^\mu \bar{\partial}^{\bar{\nu}} e_{\mu \bar{\nu}} \partial^\rho \bar{\partial}^{\bar{\sigma}} e_{\rho \bar{\sigma}}\right).
\label{highdcquad}
\end{aligned}
\end{equation}
According to \cite{lescan}, this action can be interpreted as a higher-derivative extension of DFT with conformal symmetry in the double
space up to quadratic order. The authors clarify this point by showing that in the supergravity frame and properly rescaling the metric and/or considering a particular volume for the double space for integration, the action (\ref{highdcquad}) reduces to the conformal gravity action with a gauge fixing. \\

The next step is to extend this procedure to cubic order. For the Lagrangian (\ref{higherymlag}), the action up to cubic order takes the form 
\begin{equation}
\begin{aligned}
    S_{\mathrm{HD}}^{(3)}=g_{\mathrm{YM}} \int d^D x f_{a b c}\bigg(\triangle_\nu{ }^a&\left(A_\rho{ }^b \partial^\rho A^{\nu c}-A_\rho{ }^b \partial^\nu A^{\rho c}\right)\\
    &-\partial_\rho \triangle_\nu{ }^a A^{\rho b} A^{\nu c}\bigg),
\end{aligned}
\end{equation}
with $\triangle_\nu{ }^a \equiv \square A_\nu{ }^a-\partial_\nu \partial^\mu A_\mu{ }^a .$ In momentum space, this action can be written as
\begin{equation}
\begin{aligned}
    S_{\mathrm{HD}}^{(3)}=\frac{i g_{\mathrm{YM}}}{(2 \pi)^{D / 2}} & \int d k_1 d k_2 d k_3 \, \delta\left(k_1+k_2+k_3\right) f_{a b c} \, k_1^2 \\
    &\times \Pi^{\rho \mu}\left(k_3^\nu-k_1^\nu\right) A_{1 \mu}{ }^a A_{2 \nu}{ }^b A_{3 \rho}{ }^c,
    \label{action29}
\end{aligned}
\end{equation}
where $\Pi^{\rho \mu}$ is defined in (\ref{projector}). Considering the deformation to a three-point vertex function given by 
\begin{equation}
    \Pi^{\mu \nu \rho}\left(k_1, k_2, k_3\right)=\Pi^{\mu \nu} k_{12}^\rho+\Pi^{\nu \rho} k_{23}^\mu+\Pi^{\rho \mu} k_{31}^\nu ,
    \label{highcubicproj}
\end{equation}
the action (\ref{action29}) can be written more symmetrically as
\begin{equation}
    S_{\mathrm{HD}}^{(3)}=\frac{i g_{\mathrm{YM}}}{3(2 \pi)^{D / 2}} \int_{k_i} \delta\left(k_i\right) f_{a b c} \, k_1^2 \, \Pi^{\mu \nu \rho} A_{1 \mu}{ }^a A_{2 \nu}{ }^b A_{3 \rho}{ }^c,
    \label{action31}
\end{equation}
where $\delta(k_i)=\delta(k_1 + k_2 + k_3)$. Notice that the definition (\ref{highcubicproj}) for the projector of the cubic theory differs from the definition (\ref{cubicproj}) in the previous section. The authors verify this point by saying that the only way to apply the double copy prescription preserving the symmetry between the coordinates $\mu$ and $\Bar{\mu}$ is to consider $\Pi^{\mu \nu \rho}$ instead of $\pi^{\mu \nu \rho}$ to replace the structure constant as
\begin{equation}
    f_{a b c} \longrightarrow \frac{i}{4} \bar{\Pi}^{\bar{\mu} \bar{\nu} \bar{\rho}} .
\end{equation}
Then, imposing the double copy prescription on (\ref{action31}),
\begin{equation}
\begin{aligned}
    S_{\mathrm{HDC}}^{(3)}=-\frac{1}{24(2 \pi)^{D / 2}} &\int d K_1 d K_2 d K_3 \, \delta\left(K_1+K_2+K_3\right)\\ &\times (k_1)^2 \bar{\Pi}^{\bar{\mu} \bar{\nu} \bar{\rho}} \, \Pi^{ \mu \nu \rho} \, e_{1 \mu \bar{\mu}} e_{2 \nu \bar{\nu}} e_{3 \rho \bar{\rho}} .
    \label{eqn33}
\end{aligned}
\end{equation}
Expanding this action by using the definitions of projectors gives rise to non-local terms once transformed
to coordinate space. Introducing an auxiliary scalar field and transforming to the coordinate space, the final form has been found as \cite{lescan}
\begin{widetext}
\begin{equation}
    \begin{aligned}
S_{\mathrm{HDC}}^{(3)}=  - \frac{1}{4} \int_{x, \bar{x}}&\left(  \square \partial^\rho \bar{\partial}^{\bar{\sigma}} e_{\mu \bar{\nu}} e^{\mu \bar{\nu}} e_{\rho \bar{\sigma}}-\square \bar{\partial}^{\bar{\sigma}} e_{\mu \bar{\nu}} \partial^\rho e^{\mu \bar{\nu}} e_{\rho \bar{\sigma}}+\square \bar{\partial}^{\bar{\sigma}} e_{\mu \bar{\nu}} \partial^\mu e^{\rho \bar{\nu}} e_{\rho \bar{\sigma}}-\square \bar{\partial}^{\bar{\sigma}} e_{\mu \bar{\nu}} e^{\rho \bar{\nu}} \partial^\mu e_{\rho \bar{\sigma}}\right. \\
& -\square e_{\mu \bar{\nu}} \partial^\mu \bar{\partial}^{\bar{\sigma}} e^{\rho \bar{\nu}} e_{\rho \bar{\sigma}}+\square e_{\mu \bar{\nu}} \bar{\partial}^{\bar{\sigma}} e^{\rho \bar{\nu}} \partial^\mu e_{\rho \bar{\sigma}}-\partial^\mu \partial^\rho \partial^\lambda \bar{\partial}^{\bar{\sigma}} e_{\mu \bar{\nu}} e_\rho{}^{\bar{\nu}} e_{\lambda \bar{\sigma}}+\partial^\mu \partial^\rho \bar{\partial}^{\bar{\sigma}} e_{\mu \bar{\nu}} \partial^\lambda e_\rho{}^{\bar{\nu}} e_{\lambda \bar{\sigma}} \\
& -\partial^\rho \partial^\lambda \bar{\partial}^{\bar{\sigma}} e_{\mu \bar{\nu}} \partial^\mu e_\rho{}^{\bar{\nu}} e_{\lambda \bar{\sigma}}+\partial^\rho \partial^\lambda \bar{\partial}_{\bar{\sigma}} e_{\mu \bar{\nu}} e_\rho{}^{\bar{\nu}} \partial^\mu e_{\lambda \bar{\sigma}}+\partial^\rho \partial^\lambda e_{\mu \bar{\nu}} \partial^\mu \bar{\partial}^{\bar{\sigma}} e_\rho{}^{\bar{\nu}} e_{\lambda \bar{\sigma}}-\partial^\rho \partial^\lambda e_{\mu \bar{\nu}} \bar{\partial}_{\bar{\sigma}} e_\rho{}^ {\bar{\nu}} \partial^\mu e_{\lambda \bar{\sigma}} \\
& -\bar{\partial}^{\bar{\nu}} \bar{\partial}^{\bar{\sigma}} \bar{\partial}^{\bar{\kappa}} \partial^\rho e_{\mu \bar{\nu}} e^\mu{}_{\bar{\sigma}} e_{\rho \bar{\kappa}}+\bar{\partial}^{\bar{\nu}} \bar{\partial}^{\bar{\sigma}} \bar{\partial}^{\bar{\kappa}} e_{\mu \bar{\nu}} \partial^\rho e^\mu{ }_{\bar{\sigma}} e_{\rho \bar{\kappa}}+\bar{\partial}^{\bar{\nu}} \bar{\partial}^{\bar{\sigma}} e_{\mu \bar{\nu}} \bar{\partial}^{\bar{\kappa}} \partial^\mu e^\rho{ }_{\bar{\sigma}} e_{\rho \bar{\kappa}}-\bar{\partial}^{\bar{\nu}} \bar{\partial}^{\bar{\sigma}} e_{\mu \bar{\nu}} \bar{\partial}^{\bar{\kappa}} e^\rho{ }_{\bar{\sigma}} \partial^\mu e_{\rho \bar{\kappa}} \\
& \left.+\partial^\mu \partial^\rho \bar{\partial}^{\bar{\nu}} \bar{\partial}^{\bar{\sigma}} \phi e_{\mu \bar{\nu}} e_{\rho \bar{\sigma}}-\partial^\mu \bar{\partial}^{\bar{\nu}} \bar{\partial}^{\bar{\sigma}} \phi \partial^\rho e_{\mu \bar{\nu}} e_{\rho \bar{\sigma}}+\partial^\mu \partial^\rho \phi \bar{\partial}^{\bar{\nu}} \bar{\partial}^{\bar{\sigma}} e_{\mu \bar{\nu}} e_{\rho \bar{\sigma}}-\partial^\rho \phi \bar{\partial}^{\bar{\nu}} \bar{\partial}^{\bar{\sigma}} e_{\mu \bar{\nu}} \partial^\mu e_{\rho \bar{\sigma}}\right).
\end{aligned}
\label{lescanofinal}
\end{equation}
\end{widetext}
In \cite{lescan}, the authors state that the cubic action obtained from the double copy of the $(DF)^2$ theory contains further contributions apart from the Weyl-square action when setting $x=\Bar{x}$ for the pure gravity case $\left(e_{\mu \bar{\nu}} \sim h_{\mu \nu}, \phi=0\right)$. Some of these extra terms vanish for a particular gauge choice and a particular spacetime dimension, but the result still does not totally agree with the Weyl-squared action.

In \cite{lescannew}, this analysis is extended by including a charged scalar field $\phi_{\alpha}$ to the Lagrangian. The authors of \cite{lescannew} were inspired by \cite{johansson} and showed that the inclusion of the charged scalar field is crucial for constructing both the quadratic and cubic contributions of the potential candidate for the conformal DFT they want to construct. However, these details are not important to check the consistency of the method used in this work and so I continue to work with action (\ref{highdcquad}).\\

\section{Background independent formulation of DFT}
In the previous section, I reviewed the double copy prescription for the $(DF)^2$ extension of the Yang-Mills theory. As the authors discussed in \cite{lescan}, the resulting DFT action can be a possible candidate for a conformally invariant DFT action. In \cite{lescan}, the double copy prescription is applied up to cubic order and around a flat spacetime background. Before I attempt to extend this idea up to all orders by using manifestly background-independent objects, let me first review the background-independent formulation of DFT. 
In \cite{dft}, it has been shown that the DFT action for the dilaton $d$ and the fluctuation field $e_{ij}$ around the constant background $E_{ij}$ up to cubic order is 
\begin{widetext}
\begin{equation}
    \begin{aligned}
S= & \int(d x d \tilde{x})\left(\frac{1}{4} e_{i j} \square e^{i j}+\frac{1}{4}\left(\bar{D}^j e_{i j}\right)^2+\frac{1}{4}\left(D^i e_{i j}\right)^2-2 d D^i \bar{D}^j e_{i j}-4 d \square d \right. \\
&+\frac{1}{4} e_{i j}\left(\left(D^i e_{k l}\right)\left(\bar{D}^j e^{k l}\right)-\left(D^i e_{k l}\right)\left(\bar{D}^l e^{k j}\right)-\left(D^k e^{i l}\right)\left(\bar{D}^j e_{k l}\right)\right) \\
&\left. +\frac{1}{2} d\left(\left(D^i e_{i j}\right)^2+\left(\bar{D}^j e_{i j}\right)^2+\frac{1}{2}\left(D_k e_{i j}\right)^2+\frac{1}{2}\left(\bar{D}_k e_{i j}\right)^2+2 e^{i j}\left(D_i D^k e_{k j}+\bar{D}_j \bar{D}^k e_{i k}\right)\right)+4 e_{i j} d D^i \bar{D}^j d+4 d^2 \square d\right),
\end{aligned}
\label{dftaction}
\end{equation}
\end{widetext}
where the derivative operators are defined by
\begin{equation}
\begin{aligned}
&D_i \equiv \frac{\partial}{\partial x^i}-E_{i k} \frac{\partial}{\partial \tilde{x}_k}, \quad \quad \bar{D}_i\equiv \frac{\partial}{\partial x^i}+E_{i k}^t \frac{\partial}{\partial \tilde{x}_k},\\
&\square \equiv \frac{1}{2}\left(D^2+\bar{D}^2\right).
\label{derivop}
\end{aligned}
\end{equation}
In \cite{hohmback}, the action (\ref{dftaction}) is shown to be background-independent to quadratic order. To write this action in a manifestly background-independent way, a new object is introduced as
\begin{equation}
    \mathcal{E}_{i j} \equiv E_{i j}+e_{i j}+\frac{1}{2} e_i^k e_{k j}+\mathcal{O}\left(e^3\right),
    \label{backindepfield}
\end{equation}
which combines the background and the fluctuation fields. The action (\ref{dftaction}) can be written in terms of this new object, by modifying the definitions of the derivative operators in (\ref{derivop}) as
\begin{equation}
\mathcal{D}_i  \equiv \partial_i-\mathcal{E}_{i k} \tilde{\partial}^k \quad \text{and} \quad \bar{\mathcal{D}}_i  \equiv \partial_i+\mathcal{E}_{k i} \tilde{\partial}^k.
\label{backindepderiv}
\end{equation}
 The calligraphic derivatives (\ref{backindepderiv}) can be written in terms of (\ref{derivop}) as
\begin{equation}
    \begin{aligned}
\mathcal{D}_i & =D_i-\frac{1}{2} e_{i k}(\bar{D}^k-D^k)-\frac{1}{4} e_i{}^k e_{k l}(\bar{D}^l-D^l)+\mathcal{O}(e^3), \\
\bar{\mathcal{D}}_i & =\bar{D}_i+\frac{1}{2} e_{k i}(\bar{D}^k-D^k)+\frac{1}{4} e_k{}^l e_{l i}(\bar{D}^k-D^k)+\mathcal{O}(e^3).
\end{aligned}
\label{backindepderivfinal}
\end{equation}
 The indices of these background-independent derivatives can be raised by the full metric so that
\begin{equation}
    \begin{aligned}
& \mathcal{D}^i \equiv g^{i j} \mathcal{D}_j=D^i-\frac{1}{2} e^{i j} \bar{D}_j-\frac{1}{2} e^{j i} D_j+\mathcal{O}\left(e^2\right), \\
& \bar{\mathcal{D}}^i \equiv g^{i j} \bar{\mathcal{D}}_j=\bar{D}^i-\frac{1}{2} e^{k i} D_k-\frac{1}{2} e^{i k} \bar{D}_k+\mathcal{O}\left(e^2\right),
\end{aligned}
\end{equation}
 and $g^{ij}$, the inverse of the full metric $g_{i j}=\frac{1}{2}\left(\mathcal{E}_{i j}+\mathcal{E}_{j i}\right)$, reads
\begin{equation}
        g^{i j}=G^{i j}-e^{(i j)}+\frac{1}{4} e^{i k} e^j{ }_k+\frac{1}{4} e^{k i} e_k{ }^j+\mathcal{O}\left(e^3\right).
\label{inversemetric}
\end{equation}
Now the question is how the action (\ref{dftaction}) can be written by using (\ref{backindepfield}) and (\ref{backindepderiv}). To achieve this, the authors of \cite{hohmback} start with the quadratic terms of (\ref{dftaction}) and consider some possible combinations of (\ref{backindepfield}) and (\ref{backindepderiv}) which can create these quadratic terms. For instance, the first term in (\ref{dftaction}) can be obtained from 
\begin{equation}
    -\frac{1}{4} e^{-2 d} g^{i k} g^{j l} \mathcal{D}^p \mathcal{E}_{k l} \mathcal{D}_p \mathcal{E}_{i j}=
    \frac{1}{4} e^{i j} \square e_{i j}+\cdots
\end{equation}
Analogously, the ansatz for the action in terms of the background independent objects is found as \cite{hohmback}
\begin{equation}
\begin{aligned}
S=\int_{x, \bar{x}} &e^{-2 d}\bigg( -\frac{1}{4} g^{i k} g^{j l} \mathcal{D}^p \mathcal{E}_{k l} \mathcal{D}_p \mathcal{E}_{i j}+\frac{1}{4} g^{k l}(\mathcal{D}^j \mathcal{E}_{i k} \mathcal{D}^i \mathcal{E}_{j l}) \\
& +\frac{1}{4}g^{kl}(\bar{\mathcal{D}}^j \mathcal{E}_{k i} \bar{\mathcal{D}}^i \mathcal{E}_{l j})+(\mathcal{D}^i d \bar{\mathcal{D}}^j \mathcal{E}_{i j}+\bar{\mathcal{D}}^i d \mathcal{D}^j \mathcal{E}_{j i})\\
&+4 \mathcal{D}^i d \mathcal{D}_i d\bigg).
\end{aligned}
\label{backindepaction}
\end{equation}
This manifestly background-independent DFT action reproduces all the quadratic terms in (\ref{dftaction}) by construction. Remarkably, when this action is expanded up to cubic orders, it also exactly matches with the cubic terms in (\ref{dftaction}). Thus, by starting from the quadratic part of the DFT action, it is possible to construct a manifestly background-independent DFT action up to higher orders. This action is background independent when the objects are expanded up to all orders.
\section{Higher order extensions of higher derivative DFT}
In this section, I offer a method to directly obtain third-order terms (\ref{lescanofinal}) from the quadratic terms (\ref{highdcquad}) by using the idea introduced in Section 4. The first observation is that the action (\ref{backindepaction}) matches with (\ref{dcquad}) when the terms expanded up to first order and the background field is taken as $\eta_{ij}$, namely
\begin{equation}
    \mathcal{D}_i = D_i, \quad \bar{\mathcal{D}}_i = \bar{D}_i, \quad \mathcal{E}_{ij} = \eta_{ij} + e_{ij}, \quad g_{ij}=\eta_{ij},
\label{def1}
\end{equation}
and
\begin{equation}
    d= -\frac{1}{4} \phi.
\end{equation}
Similarly, when the terms in (\ref{backindepaction}) expanded up to second order, namely,
\begin{equation}
\begin{aligned}
\mathcal{D}_i= &D_i-\frac{1}{2} e_{i k}(\bar{D}^k-D^k), \quad \bar{\mathcal{D}}_i =\bar{D}_i+\frac{1}{2} e_{k i}(\bar{D}^k-D^k) \\
\mathcal{E}_{ij}= & \eta_{i j}+e_{i j}+\frac{1}{2} e_i{}^k e_{k j}, \quad d= -\frac{1}{4} \phi',
\end{aligned}
\label{definitions}
\end{equation}
where $\phi'$ refers to the second order expansion of the dilaton $\phi$, the action (\ref{backindepaction}) matches with (\ref{dccubic}) after a gauge is chosen.\\

So we can ask if it is possible to use the background independent objects to define an action that gives (\ref{highdcquad}) when it is expanded up to second order in fields and gives (\ref{lescanofinal}) when it is expanded up to third order in fields. My starting point is the quadratic action (\ref{highdcquad}). A possible candidate which can create the first term of (\ref{highdcquad}) is  
 \begin{equation}
     -\frac{1}{4}e^{-2d} g'^{\mu k} g'^{\nu \tau} g'^{\alpha m} g'^{\beta n}\left(\mathcal{D}' _\alpha \mathcal{D}'_m \mathcal{E}'_{k \tau}\right)\left({\mathcal{D}}'_\beta {\mathcal{D}}'_n \mathcal{E}'_{\mu \nu}\right),
    \label{ansatz11}
 \end{equation}
where the primes on the terms imply that these objects are defined only up to the second orders in their expansions. This means that $\mathcal{E}_{ij}$ is defined up to second order in $e_{ij}$, where $g_{ij}$ and $\mathcal{D}_{ij}$ are defined up to first order in $e_{ij}$. However, as discussed in \cite{hohmback}, more than one derivative or no derivatives on $\mathcal{E}_{ij}$ would imply complications with the $O(D,D)$ covariance because of the contraction of barred and unbarred indices in the definitions (\ref{backindepfield}) and (\ref{backindepderiv}). On the other hand, the terms with one derivative on $\mathcal{E}_{ij}$ transform as a tensor under $O(D,D)$. Therefore, I should be careful when I introduce the higher derivative terms. Following \cite{hohmback}, let me first introduce the Christoffel-like connections:
\begin{equation}
   \begin{aligned} \Gamma_{i \bar{j}}^{\bar{k}} & \equiv \frac{1}{2} g^{k l}\left(\mathcal{D}_i \mathcal{E}_{l j}+\bar{\mathcal{D}}_j \mathcal{E}_{i l}-\bar{\mathcal{D}}_l \mathcal{E}_{i j}\right) \\ \Gamma_{\overline{i}{j}}^k & \equiv \frac{1}{2} g^{k l}\left(\bar{\mathcal{D}}_i \mathcal{E}_{j l}+\mathcal{D}_j \mathcal{E}_{l i}-\mathcal{D}_l \mathcal{E}_{j i}\right),\end{aligned}
\label{connections}
\end{equation}
and $O(D,D)$ covariant derivatives
\begin{equation}
    \nabla_i \bar{\eta}_j  \equiv \mathcal{D}_i \bar{\eta}_j-\Gamma_{i \bar{j}}^k \bar{\eta}_k, \quad \bar{\nabla}_j \eta_i\equiv \overline{\mathcal{D}}_j \eta_i-\Gamma_{\bar{j} i}^k \eta_k.
    \label{covderiv}
\end{equation}
Similarly, for the remaining index configurations
\begin{equation}
    \begin{array}{ll}\nabla_i \eta_j=\mathcal{D}_i \eta_j-\Gamma_{i j}^k \eta_k, & \quad \Gamma_{i j}^k=\frac{1}{2} g^{k l} \mathcal{D}_i \mathcal{E}_{j l}, \\\\ \bar{\nabla}_i \bar{\eta}_j=\bar{\mathcal{D}}_i \bar{\eta}_j-\Gamma_{\bar{i} \bar{j}}^{\bar{k}} \bar{\eta}_k, & \quad \Gamma_{\bar{i} \bar{j}}^{\bar{k}}=\frac{1}{2} g^{k l} \bar{\mathcal{D}}_i \mathcal{E}_{l j}.\end{array}
    \label{remainindex}
\end{equation}
Note that these covariant derivatives are metric-compatible
\begin{equation}
    \nabla_i g_{j k}=0, \quad \quad \bar{\nabla}_i g_{j k}=0,
\end{equation}
where the indices of the metric can be both barred or unbarred.

Now I can write my ansatz for the first term (\ref{ansatz11})  in an $O(D,D)$ covariant way as
\begin{equation}
    -\frac{1}{4} e^{-2d'} g'^{\mu k} g'^{\nu \tau} g'^{\alpha m} g'^{\rho q}\left(\nabla'_\alpha \mathcal{D}'_\rho \mathcal{E}'_{k \tau}\right)\left({\nabla}'_m {\mathcal{D}}'_q \mathcal{E}'_{\mu \nu}\right),
    \label{ansatz1final}
\end{equation}
where the prime on the $\nabla$ operator indicates that it is defined up to the first order in $e_{ij}$. This action is, up to second order in fields, equal to
\begin{equation}
    \mathcal{A}_1^{(2)} = -\frac{1}{4} (D_\alpha D_\rho e^{\mu \nu})({D}^\alpha {D}^\rho e_{\mu \nu}).
\end{equation} 
To expand (\ref{ansatz1final}) to higher orders, I substituted the definitions (\ref{backindepfield}), (\ref{backindepderivfinal}), (\ref{inversemetric}), (\ref{connections}), (\ref{covderiv}), (\ref{remainindex}) into (\ref{ansatz1final}). After a long but straightforward calculation (AppendixA), the terms up to cubic order are found as
\begin{equation}
    \begin{aligned}
\mathcal{A}_1^{(3)}=& +\frac{1}{4}(D^\alpha {D}^\rho e^{\mu \nu})(D_\rho e^\mu{}_p)\left(\bar{D}^\nu e_\alpha{}^p - \bar{D}^p e_\alpha{}^\nu \right) \\
&+\frac{1}{4}(D^\alpha D^\rho e^{\mu \nu}) \left( e_{\alpha p} (\bar{D}^p D_\rho e_{\mu \nu}) + 2d (D_\alpha D_\rho e_{\mu \nu})\right) \\
& -\frac{1}{4}e_{\alpha p}(\bar{D}^p e_{\mu \nu})(\square D^\alpha e^{\mu \nu}).
\label{mycubic1}
\end{aligned}
\end{equation}
Now, I move to the ansatz for the second term in (\ref{highdcquad}). A possible candidate to create this term is
\begin{equation}
    \mathcal{A}_2 \equiv \frac{1}{4}e^{-2d'} g'^{\mu k} g'^{\nu \tau} g'^{\rho \omega} g'^{\alpha m}\left(\nabla'_\alpha \mathcal{D}'_\omega \mathcal{E}'_{k \tau}\right)\left(\nabla'_m \mathcal{D}'_\mu \mathcal{E}'_{\rho \nu}\right),
    \label{ansatz2}
\end{equation}
which can be expanded as
\begin{equation}
\begin{aligned}
\mathcal{A}_2^{(2)} = &+\frac{1}{4}\left(D_\alpha {D}^\rho e^{\mu \nu}\right)\left(D^\alpha {D}_\mu {e}_{\rho \nu}\right),\\
\mathcal{A}_2^{(3)} =&+\frac{1}{4}(D^\alpha D_\mu e_{\rho \nu})(D^\rho e^\mu{}_p)\left(\bar{D}^p e_\alpha{}^\nu - \bar{D}^\nu e_\alpha{}^p\right)\\
&-\frac{1}{4}({D}^\alpha {D}_\mu e_{\rho \nu}) \left(e_{\alpha p}(D^\rho \bar{D}^p e^{\mu \nu}) + 2d (D_\alpha D^\rho e^{\mu \nu})\right) \\
&+\frac{1}{4}e^\alpha{}_p(\bar{D}^p e^{\mu \nu})(\square {D}_\mu e_{\alpha \nu}).
\end{aligned}
\label{ansatz2third}
\end{equation}
Similarly, an ansatz for the third term in (\ref{highdcquad}) takes the form
\begin{equation}
    \mathcal{A}_3\equiv \frac{1}{4} e^{-2d'} g'^{\mu k} g'^{\nu \tau} g'^{\sigma \omega} g'^{p m}\left(\bar{\nabla}_p \bar{\mathcal{D}}'_\omega \mathcal{E}'_{k \tau}\right)\left(\bar{\nabla}'_m \bar{\mathcal{D}}'_\nu \mathcal{E}'_{\mu \sigma}\right).
\label{ansatz3}
\end{equation}
This term gives up to second and third order,
\begin{equation}
     \begin{aligned}
\mathcal{A}_3^{(2)} = &+\frac{1}{4} (\bar{D}_p \bar{{D}}^\sigma {e}_{\mu \nu})\left(\bar{D}^p \bar{{D}}_\nu {e}_{\mu \sigma}\right), \\
\mathcal{A}_3^{(3)} =&+\frac{1}{4}(\bar{D}^p \bar{D}_\nu e_{\mu \sigma})(\bar{D}^\sigma e_\alpha{}^\nu)\left(D^\alpha e^\mu{}_p - D^\mu e^\alpha{}_p \right)\\
&-\frac{1}{4}(\bar{D}^p \bar{D}_\nu e_{\mu \sigma}) \left(e_{\alpha p}(D^\alpha \bar{D}^\sigma e^{\mu \nu}) +2d(\bar{D}_p \bar{D}^\sigma e^{\mu \nu})\right) \\
&+\frac{1}{4}e_\alpha{}^p(D^\alpha e^{\mu \nu})(\square \bar{D}_\nu e_{\mu p}).
\end{aligned}
\end{equation} 
For the last term in (\ref{highdcquad}), I can employ an ansatz as
\begin{equation}
     \mathcal{A}_4 \equiv -\frac{1}{4}e^{-2d'} g'^{\mu k} g'^{\nu \tau} g'^{\alpha \sigma} g'^{\beta \rho}\left(\nabla'_k \bar{\mathcal{D}}'_\tau \mathcal{E}'_{\mu \nu}\right)\left(\bar{\nabla}'_\alpha {\mathcal{D}}'_\beta \mathcal{E}'_{\rho \sigma}\right).
     \label{ansatz4new}
\end{equation}
However, instead of using an ansatz for the last term of (\ref{highdcquad}) directly, first I introduce the dilaton term in (\ref{highdcquadmom}) and write it as
\begin{equation}
\begin{aligned}
    S_{\mathrm{HDC}}^{(2)}=-\frac{1}{4} \int_{k, \bar{k}} k^2(&k^2 e^{\mu \bar{\nu}} e_{\mu \bar{\nu}}-k^\mu k^\rho e_{\mu \bar{\nu}} e_\rho^{\bar{\nu}}-\bar{k}^{\bar{\nu}} \bar{k}^{\bar{\sigma}} e_{\mu \bar{\nu}} e^\mu{ }_{\bar{\sigma}} \\
    & +\phi k^\mu \bar{k}^{\bar{\nu}} e_{\mu \bar{\nu}} ),
\end{aligned}
\end{equation}
which takes the form
\begin{equation}
\begin{aligned}
S_{\mathrm{HDC}}^{(2)}=-\frac{1}{4} &\int d^D x d^D \bar{x} \left(\square e^{\mu \bar{\nu}} \square e_{\mu \bar{\nu}}-\square e^{\mu \bar{\nu}} \partial_\mu \partial^\rho e_{\rho \bar{\nu}} \right.\\
& \left. -\square e^{\mu \bar{\nu}} \bar{\partial}_{\bar{\nu}} \bar{\partial}^{\bar{\sigma}} e_{\mu \bar{\sigma}}+\square \phi \partial^\mu \bar{\partial}^{\bar{\nu}} e_{\mu \bar{\nu}} \right),
\end{aligned}
\label{highdcquadnew}
\end{equation}
in the position space. Now, I can use an ansatz for the last term of (\ref{highdcquadnew}) by
\begin{equation}
    \mathcal{A}_{4}\equiv -\frac{1}{4}g'^{\mu k} g'^{\nu \tau} g'^{\alpha m}(\nabla'_\alpha \mathcal{D}'_\mu \phi)(\nabla'_m \bar{\mathcal{D}}'_\nu \mathcal{E}'_{k \tau}),
\end{equation}
which can be expanded as
\begin{equation}
\begin{aligned}
\mathcal{A}_4^{(2)} = &-\frac{1}{4} ({D}_\alpha {D}_\mu \phi)({D}^\alpha \bar{{D}}_\nu {e}^{\mu \nu}), \\
\mathcal{A}_4^{(3)} =&+\frac{1}{8}e_{\alpha p}(\bar{D}^p {D}_\mu \phi)({D}^\alpha \bar{D}_\nu e^{\mu \nu})\\
&+\frac{1}{8}(D^\alpha {D}_\mu \phi) \left( e_{\alpha p}(\bar{D}^p \bar{D}_\nu e^{\mu \nu}) + 4d(D_\alpha \bar{D}_\nu e^{\mu \nu}) \right) \\
&-\frac{1}{8}e_{\mu p}(\bar{D}^p \phi)(\square \bar{D}_\nu e^{\mu \nu}) \\
&-\frac{1}{8}e_{\alpha p}(D^\alpha e^{\mu \nu})(\square D_\mu \phi).
\end{aligned}
\end{equation} 
To recapitulate, I have started with the action (\ref{highdcquad}) introduced in \cite{lescan} and attempted to write this action by using the background independent objects up to the third order in fields. At this point, it is important to note that the resulting action is not background independent because the expansions of the manifestly background independent objects are taken only up to their second orders. The background independence requires the expansions of these objects up to all orders. In the end, by combining all of the ansatz terms introduced above, my final result reads
\begin{equation}
    \begin{aligned}
 \mathcal{L}_2=e^{-2d'} \bigg( & -\frac{1}{4} g'^{u k} g'^{\nu \tau} g'^{\alpha m} g'^{\rho q}(\nabla'_\alpha \mathcal{D}'_\rho \mathcal{E}'_{k \tau})({\nabla}'_m {\mathcal{D}}'_q \mathcal{E}'_{\mu \nu}) \\
& +\frac{1}{4} g'^{\mu k} g'^{\nu \tau} g'^{\rho \omega} g'^{\alpha m}(\nabla'_\alpha \mathcal{D}'_\omega \mathcal{E}'_{k \tau})(\nabla'_m \mathcal{D}'_\mu \mathcal{E}'_{\rho \nu})\\
&+\frac{1}{4} g'^{\mu k} g'^{\nu \tau} g'^{\sigma \omega} g'^{\alpha m}(\bar{\nabla}'_\alpha \bar{\mathcal{D}}'_\omega \mathcal{E}'_{k \tau})(\bar{\nabla}'_m \bar{\mathcal{D}}'_\nu \mathcal{E}'_{\mu \sigma})\\
& -\frac{1}{4} g'^{\mu k}g'^{\nu \tau}g'^{\alpha m} ( \nabla'_\alpha {\mathcal{D}}'_\mu \phi) ( {\nabla}'_m \bar{\mathcal{D}}'_\nu \mathcal{E}'_{k \tau}) \bigg)
\end{aligned}
\label{ansatzfinal}
\end{equation}
which gives up to second order
\begin{equation}
    \begin{aligned}
 \mathcal{L}_2^{(2)}=&-\frac{1}{4}\left(\square e^{\mu \nu}\right)\left(\square e_{\mu \nu}\right)-\frac{1}{4}\left({D}_\mu \bar{{D}}_\nu {e}^{\mu \nu}\right)\left(\square \phi\right)\\
 &+\frac{1}{4}\left(\square e^{\mu \nu}\right)\left(\bar{D}_\nu \bar{D}^\sigma e_{\mu \sigma}\right)+\frac{1}{4}\left(\square e^{\mu \nu}\right)\left(D_\mu D^\rho e_{\rho \nu}\right),
\end{aligned}
\label{mysecondorder}
\end{equation}
and up to the third order
\begin{equation}
    \mathcal{L}_2^{(3)}= \mathcal{L}^{(3)}_{21} + \mathcal{L}^{(3)}_{22} + \mathcal{L}^{(3)}_{23}+ \mathcal{L}^{(3)}_{24}+ \mathcal{L}^{(3)}_{25}+ \mathcal{L}^{(3)}_{26}
\end{equation}
with
\begin{widetext}
\begin{align}
\mathcal{L}^{(3)}_{21}  \equiv &-\frac{1}{4} e_{\rho \sigma}\left(\bar{D}^\sigma e_{\mu \nu}\right)\left(D^\rho \square e^{\mu \nu}\right)+\frac{1}{4} e_{\rho \sigma}\left(D^\rho D^\alpha e_{\mu \nu}\right)\left(\bar{D}^\sigma D_\alpha e^{\mu \nu}\right), \nonumber \\
\mathcal{L}^{(3)}_{22} \equiv &+\frac{1}{4}(\bar{D}^\sigma e_\mu{}^\nu)\bigg[(D^\mu D^\alpha e_{\rho \sigma})(D_\alpha e^\rho{}_\nu)-(D^\mu D^\alpha e_{\rho \nu})(D_\alpha e^\rho{}_\sigma)\bigg]+\frac{1}{4}e^\rho{}_\sigma(\square D_\mu e_{\rho \nu})(\bar{D}^\sigma e^{\mu \nu})+\frac{1}{4}e_\mu{}^\nu (\square \bar{D}_\sigma e_{\rho \nu})(D^\mu e^{\rho \sigma}), \nonumber \\
\mathcal{L}^{(3)}_{23} \equiv & +\frac{1}{4}(D^\mu D^\lambda e_{\rho \nu})(\bar{D}^\sigma e_{\mu}{}^\nu)(D^\rho e_{\lambda \sigma})-\frac{1}{4}(D^\rho D^\mu e_{\lambda \sigma})(\bar{D}^\sigma e_{\rho}{}^\nu)(D^\lambda e_{\mu \nu})-\frac{1}{4}e_{\lambda \sigma}(D^\sigma D^\rho e_{\mu}{}^\nu)(D^\lambda D^\mu e_{\rho \nu}), \nonumber \\
\mathcal{L}^{(3)}_{24} \equiv & +\frac{1}{4}(\bar{D}^\kappa \bar{D}^\nu e^\mu{}_\sigma)(D^\rho e_{\mu \nu})(\bar{D}^\sigma e_{\rho \kappa})-\frac{1}{4}(\bar{D}^\sigma \bar{D}^\nu e_{\rho \kappa})(D^\rho e^{\mu}{}_\sigma)(\bar{D}^\kappa e_{\mu}{}^\nu)-\frac{1}{4}e_{\rho \kappa}(D^\rho \bar{D}^\sigma e_{\mu \nu})(\bar{D}^\kappa \bar{D}^\nu e^{\mu}{}\sigma), \nonumber \\
\mathcal{L}^{(3)}_{25} \equiv & + \frac{1}{8}e_{\alpha p}\bigg[(\bar{D}^p D_\mu \phi)(D^\alpha \bar{D}_\nu e^{\mu \nu}) + (D^\alpha D_\mu \phi)(\bar{D}^p \bar{D}_\nu e^{\mu \nu})\bigg]-\frac{1}{8}e_{\mu p}(\bar{D}^p \phi)(\square \bar{D}_\nu e^{\mu \nu})-\frac{1}{8}e_{\alpha p}(D^\alpha e^{\mu \nu})(\square D_\mu \phi), \nonumber \\
\mathcal{L}^{(3)}_{26} \equiv & -\frac{1}{8} \phi \left((D^\alpha D^\rho e^{\mu \nu})(D_\alpha D_\rho e_{\mu \nu})-(D^\alpha D_\mu e_{\rho \nu})(D_\alpha D^\rho e^{\mu \nu})-(\bar{D}^p \bar{D}_\nu e_{\mu \sigma})(\bar{D}^p \bar{D}^\sigma e^{\mu \nu})+(D^\alpha D_{\mu} \phi)(D_\alpha \bar{D}_\nu e^{\mu \nu}) \right),
\end{align}
\end{widetext}
where I used index manipulations and integration by parts to write the result in a compact form. I classified and separated third order terms into six parts to make the comparison with (\ref{lescanofinal}) easier. For instance, I expect the $\mathcal{L}^{(3)}_{21}$ terms to match with the first two terms of (\ref{lescanofinal}). Similarly, $\mathcal{L}_{25}^{(3)}$ and $\mathcal{L}_{25}^{(3)}$ parts should correspond to the dilaton terms of (\ref{lescanofinal}). However, my analysis shows that the cubic terms coming from (\ref{ansatzfinal}) do not match with (\ref{lescanofinal}).

To compare $\mathcal{L}^{(3)}_{21}$ with the first two terms of (\ref{lescanofinal}), I use integration by parts and obtain 
\begin{equation}
\begin{aligned}
    \mathcal{L}^{(3)}_{21}  \equiv & +\frac{1}{4}e_{\rho \sigma}\left((\square D^\rho \bar{D}^\sigma e^{\mu \nu})e_{\mu \nu} - (\square \bar{D}^\sigma e_{\mu \nu})(D^\rho e_{\mu \nu}) \right) \\
    &+\frac{1}{4}(\square {D}^\rho e_{\mu \nu})(\bar{D}^\sigma e_{\rho \sigma})e^{\mu \nu} \\
    &-\frac{1}{4}(D^\alpha e_{\rho \sigma})(D^\rho e_{\mu \nu})(\bar{D}^\sigma D_\alpha e^{\mu \nu}).
\end{aligned}
\label{mylast1integ}
\end{equation}
Although the first two terms of (\ref{mylast1integ}) have the same form as the first two terms of (\ref{lescanofinal}), they come with opposite signs, and there are two extra terms. The case is not any better for the $\mathcal{L}_{22}^{(3)}$, $\mathcal{L}_{23}^{(3)}$,  $\mathcal{L}_{24}^{(3)}$,  $\mathcal{L}_{25}^{(3)}$,  $\mathcal{L}_{26}^{(3)}$  pieces. Although there are similar structures, the terms do not match those in (\ref{lescanofinal}). At this point, it is important to notice that I am working with higher derivative DFT action, so $\alpha'$ corrections are also important. To include the extra terms coming from $\alpha'$ corrections, assume that we start with the action
\begin{equation}
    S_{DFT} = \int_{x,\tilde{x}} ( \mathcal{L}_1 +a \mathcal{L}_2),
\label{69}
\end{equation}
where $\mathcal{L}_1$ is the DFT Lagrangian defined in (\ref{backindepaction}) and $\mathcal{L}_2$ is the Lagrangian defined in (\ref{ansatzfinal}). When I expand the terms in that action to their first orders, namely when I use definitions (\ref{def1}), I obtain
\begin{equation}
    S_{DFT} = \int_{x,\tilde{x}} ( \mathcal{L}^{(2)}_1 +a \mathcal{L}^{(2)}_2),
\label{finalsecondorder}
\end{equation}
where $\mathcal{L}^{(2)}_1$ is the quadratic part of the Lagrangian defined in (\ref{backindepaction}) and  $\mathcal{L}^{(2)}_2$ is the Lagrangian defined in (\ref{mysecondorder}). Similarly, when I expand the terms in (\ref{69}) up to their second orders, namely when I use (\ref{definitions}), I obtain
\begin{equation}
    S_{DFT} = \int_{x,\tilde{x}} \left( \mathcal{L}^{(2)}_1+\mathcal{L}^{(3)}_1 +a( \mathcal{L}^{(2)}_2 +\mathcal{L}^{(3)}_2) \right).
\label{fulaction}
\end{equation}
Now, I also add $\alpha'$ corrections to my objects. Together with these corrections, the total field up to second order in fluctuations takes the form \cite{Hohm:2014xsa}
\begin{equation}
    \begin{aligned}
\mathcal{E}_{i j}= & \eta_{i j}+ e_{ij} +\frac{1}{2} e_i{}^k e_{k j} \\
& +\frac{1}{4} \alpha^{\prime}\bigg[\partial^k e_i{}^l \partial_k e_{l j}-\partial^l e_i{}^k \partial_k e_{l j}-\partial^k e_j{}^l \partial_i\left(e_{k l}-e_{l k}\right) \\
& \quad \quad +\partial^k e_i{}^l \partial_j\left(e_{k l}-e_{l k}\right)-\frac{1}{2} \partial_i e^{k l} \partial_j\left(e_{k l}-e_{l k}\right)\bigg].
\end{aligned}
\label{alphadef}
\end{equation}
By using 
\begin{equation}
    {D}_i  \equiv \partial_i-{E}_{i k} \tilde{\partial}^k \quad \text{and} \quad \bar{{D}}_i  \equiv \partial_i+{E}_{k i} \tilde{\partial}^k,
\end{equation}
with $E_{ik}= \eta_{ik}$ in our case, and leaving out the $O(D,D)$ violating terms (which will be cancelled in $O(D,D)$ invariant action), (\ref{alphadef}) becomes
\begin{equation}
    \begin{aligned}
\mathcal{E}_{i j}=  \eta_{i j}+ e_{ij} +\frac{1}{16} &\alpha' \bigg[-D^l e_i{}^k (\bar{D}_k e_{lj}) + \bar{D}^k e^l{}_j (D_i e_{lk})  \\
&+D^k e_i{}^l (\bar{D}_j e_{kl}) - \frac{1}{2}D_i e_{kl}(\bar{D}_j e^{kl}) \bigg].
\end{aligned}
\end{equation}
Substituting this to (\ref{69}), it becomes up to third order in fields and up to fourth order in derivatives
\begin{equation}
    S_{DFT} = \int_{x,\tilde{x}} \left( \mathcal{L}^{(2)}_1+\mathcal{L}^{(3)}_1 +a( \mathcal{L}^{(2)}_2 +\mathcal{L}^{(3)}_2) + \alpha' \mathcal{L}_5\right),
\label{lastaction}
\end{equation}
where the new piece is defined by
\begin{equation}
    \mathcal{L}_5 \equiv \mathcal{L}_{51}+\mathcal{L}_{52}+\mathcal{L}_{53}+\mathcal{L}_{54}+\mathcal{L}_{55},
\end{equation}
with
\begin{align}
\mathcal{L}_{51}  \equiv & \frac{1}{64} \alpha' (D_i e_{kl})(\bar{D}_j e^{kl}) (-\square e^{ij} + D_\rho D^i e^{\rho j}+ \bar{D}_p \bar{D}^j e^{ip}),\nonumber \\
\mathcal{L}_{52} \equiv & \frac{1}{32} \alpha' (D^le_i{}^k \bar{D}_k e_{lj})(-\square e^{ij}+D_\rho D^i e^{\rho j}+\bar{D}_p \bar{D}^j e^{ip}),\nonumber \\
\mathcal{L}_{53} \equiv & \frac{1}{32} \alpha' (\bar{D}^ke^l{}_j {D}_i e_{lk})(\square e^{ij}-D_\rho D^i e^{\rho j}-\bar{D}_p \bar{D}^j e^{ip}), \nonumber \\
\mathcal{L}_{54} \equiv & \frac{1}{32} \alpha' (D^ke_i{}^l \bar{D}_j e_{kl})(\square e^{ij}-D_\rho D^i e^{\rho j}-\bar{D}_p \bar{D}^j e^{ip}),\nonumber \\
\mathcal{L}_{55} \equiv &  \frac{1}{8} \alpha' (D^i \bar{D}^j d) \bigg(-D^le_{i k}\bar{D}^ke_{lj} + \bar{D}^k e^l{}_j{D}_i e_{lk}, \nonumber \\
& \quad \quad \quad \quad \quad  +(D^k e_i{}^l)(\bar{D}_j e_{kl})-\frac{1}{2}(D_i e_{kl})(\bar{D}_j e^{kl})\bigg).
\end{align}
Now I can check if adding $\alpha'$ corrections to the definition of the field $\mathcal{E}_{ij}$ can help to obtain the same results with \cite{lescan}. To see this, consider the terms that are third order in fields and fourth order in derivatives in (\ref{lastaction}), which can be written as
\begin{equation}
    S^{(4,3)}_{DFT} = \int_{x,\tilde{x}} \left(a\mathcal{L}^{(3)}_2) + \alpha' \mathcal{L}_5\right).
\end{equation}
Grouping the related terms in $\mathcal{L}^{(3)}_2$ and $\mathcal{L}_5$, I see that the $\mathcal{L}^{(3)}_{21}$ and $\mathcal{L}_{51}$ terms have similar structures and can talk to each other. So, define
\begin{equation}
    \mathcal{L}^{(3)}_{f1} \equiv a\mathcal{L}^{(3)}_{21} + \alpha'  \mathcal{L}_{51},
\end{equation}
which can be written as
\begin{equation}
    \begin{aligned}
    \mathcal{L}^{(3)}_{f1} =& -\frac{1}{4}a e_{\rho \sigma}(\bar{D}^\sigma e_{\mu \nu})\left(D^\rho \square e^{\mu \nu}\right) \\
    & +\frac{1}{4}a e_{\rho \sigma}(D^\rho D^\alpha e_{\mu \nu})(\bar{D}^\sigma D_\alpha e^{\mu \nu}) \\
    &-\frac{1}{64} \alpha' (\square e^{\rho \sigma})(D_\rho e^{\mu \nu})(\bar{D}_\sigma e^{\mu \nu}) \\
    & +\frac{1}{64} \alpha' (D_\rho D^\alpha e^{\rho \sigma})(D_\alpha e_{\mu \nu})(\bar{D}_\sigma e^{\mu \nu}) \\
    &+\frac{1}{64} \alpha' (\bar{D}_\sigma \bar{D}^\beta e^{\rho \sigma})(D_\rho e_{\mu \nu})(\bar{D}_\beta e^{\mu \nu}).
    \end{aligned}
\end{equation}
These terms are expected to match the first two terms of (\ref{lescanofinal}). While choosing the constants $\alpha'$ and $a$, I should also consider the matching between the second order terms in fields. Under these conditions, I could not find a consistent way to choose the constants to make my third order terms match those in (\ref{lescanofinal}). Unfortunately, applying integration by parts on the terms also did not help. 
\section{Discussion and Conclusion}
Before I discuss the possible reasons and implications of the mismatch mentioned in the previous section, it is useful to summarize the situation once again. As shown in \cite{jaram}, the DFT action can be constructed as a double copy of the Yang-Mills action. Starting with the quadratic terms of the Yang-Mills action (\ref{yangmills}) and applying the double copy procedure gives the quadratic terms of the full DFT action (\ref{backindepaction}). Similarly, applying the double copy procedure to the cubic part of the Yang-Mills action yields the cubic DFT action subject to a gauge condition that originates from the Siegel gauge in string field theory \cite{hohmback}, and the case is more complicated for the quartic order \cite{quartic}. However, the important point for my purposes is that we have DFT action up to all orders, even if we do not have precise double copy prescriptions for all orders. It means that I can start with the Yang-Mills action (\ref{yangmills}), find the quadratic terms of the DFT action by a double copy procedure, and extend this quadratic action up to all orders by using the method introduced in section 4. The cubic terms coming from this extended action will be the same as those obtained by a direct double copy procedure from the cubic part of the Yang-Mills action. 

The case is different for the higher derivative DFT action introduced in \cite{lescan}. This higher derivative action originates from the double copy of the higher derivative Yang-Mills action, and we do not have all the orders of this DFT action. Based on the experience with the ordinary DFT action and its double copy procedure, I can start with the quadratic part of the higher derivative DFT action (\ref{highdcquad}) and extend it to higher orders by the method introduced in Section 4. I expect that the cubic order terms coming from this extended action to be the same with (\ref{lescanofinal}), which are obtained by the double copy prescription in cubic order. My analysis shows that I can obtain similar structures for the cubic terms in these two different procedures, but the results do not match. There are discrepancies in signs and coefficients, and there are also extra terms in the cubic part of (\ref{ansatzfinal}) compared to (\ref{lescanofinal}). Adding $\alpha'$ corrections also does not solve the problem. 

What can be the reasons for this mismatch or is it possible to make them match by using a specific gauge choice? First of all, I can consider making some changes in my final ansatz (\ref{ansatzfinal}) in a way that it still matches with (\ref{highdcquad}) up to the second order. However, there aren't many alternatives to consider since $O(D,D)$ invariance restricts the structures I can use in my action. Although I considered some alternatives for the ansatz, I could not find one that matches with (\ref{lescanofinal}) in cubic order. The second possibility is that there is something wrong or a missing point about the double copy prescription while passing from the higher derivative Yang-Mills theory to the higher derivative DFT action. The double copy map between Yang-Mills theory and DFT action is shown to give useful results in \cite{jaram}, and the method used in \cite{lescan} seems consistent with that work. However, there are still some question marks about applying double copy on the Lagrangian level, and this can be the reason for the mismatch. 

It is important to realize that this double copy prescription is applied to a Yang-Mills theory in the Minkowski space background and the resulting DFT action is also in the Minkowski space background. However, DFT is background independent when it is extended to all orders. So, if this double copy procedure is extended to all orders, a background independent DFT action will be obtained from a Yang-Mills theory around a Minkowski space background. Therefore, it could be wrong to expect this procedure to work up to higher orders. The double copy formulation of DFT should be considered more carefully in the perspective of background independence.

I believe that the higher derivative DFT action obtained in \cite{lescan} can be an important point in understanding the conformal symmetry in the double configuration space. It would be really helpful to find a way to extend this action to higher orders. Moreover, writing this action in terms of background independent objects can give us more insights. If this can be achieved, then this action can also be written with a generalized metric, and it may also be possible to define a Weyl tensor in the DFT formalism. Understanding conformal symmetry in double configuration space can also help us put further constraints on the geometry of DFT, and we can perhaps fix the undetermined parts of the Riemann tensor in DFT formalism. However, my work suggests that there are still some points that should be understood about the double copy prescription of a higher derivative extension of the Yang-Mills action and its relation with a higher derivative extension of the DFT action. \\

\begin{acknowledgments}
I would like to express my deepest gratitude to Özgür Sarıoğlu and Aybike Çatal-Özer for their invaluable guidance, insightful suggestions, and continuous support throughout the development of this article. Their expertise and dedication were instrumental in shaping the research and ensuring the quality of this work. This article would not have been possible without their mentorship. I am also grateful to Olaf Hohm and Eric Lescano for the valuable discussions and the motivation they gave.
\end{acknowledgments}

\appendix
\section{Third order terms of first ansatz }
Here, I show how I constructed the third order terms in expanding the first ansatz. The ansatz for the first term in (\ref{highdcquad}) is given in (\ref{ansatz1final}) as
\begin{equation}
    -\frac{1}{4} e^{-2d'} g'^{\mu k} g'^{\nu \tau} g'^{\alpha m} g'^{\rho q}\left(\nabla'_\alpha \mathcal{D}'_\rho \mathcal{E}'_{k \tau}\right)\left({\nabla}'_m {\mathcal{D}}'_q \mathcal{E}'_{\mu \nu}\right).
\label{appendix1}
\end{equation}
Now I will expand this term up to cubic order by substituting the definitions (\ref{backindepfield}), (\ref{backindepderivfinal}),(\ref{inversemetric}), (\ref{connections}), (\ref{covderiv}), (\ref{remainindex}) to (\ref{ansatz1final}). Notice that if I use only the first terms in these definitions, I find the term up to second order in the expansion
\begin{equation}
    -\frac{1}{4} \left(D_\alpha {D}_\rho {e}^{\mu \nu}\right)\left({D}^\alpha {{D}}^\rho {e}_{\mu \nu}\right),
\end{equation}
where I have also used $e^{-2d}\cong(1-2d)$ up to first order. To find the third order terms, I should consider the second terms in expanding the definitions only for one ingredient in (\ref{appendix1}) and the first for the other. For instance, a third order term can be found as
\begin{equation}
\begin{aligned}
    -\frac{1}{4} \left(G^{\mu k}-\frac{1}{2}e^{\mu k}-\frac{1}{2}e^{k \mu}\right) G^{\nu \tau} \left(D_\alpha {D}_\rho {e}_{k \tau}\right)\left({D}^\alpha {{D}}^\rho {e}_{\mu \nu}\right),
    \label{metric1exp}
\end{aligned}
\end{equation}
so I use this procedure for all the ingredients of (\ref{appendix1}). Another important point is the consistent index contractions for $O(D,D)$ covariance. For instance, in (\ref{metric1exp}), the new terms with $e^{\mu k}$ and $e^{k \mu}$ cause some contractions of barred and unbarred indices. These $O(D,D)$ symmetry breaking terms should cancel each other because I used $O(D,D)$ covariant derivatives. Now, I will start my calculation by separating the case into three parts:
\begin{enumerate}
    \item Third order terms coming from the connection parts of $\nabla_\alpha$,
    \item Third order terms coming from the connection parts of ${\nabla}_m$,
    \item Other terms.
\end{enumerate}
Let me start from
\begin{equation}
    -\frac{1}{4} G^{\mu k} G^{\nu \tau} G^{\alpha m} G^{\rho q}\left(\nabla_\alpha {D}_\rho {e}_{k \tau}\right)\left({D}_m {{D}}_q {e}_{\mu \nu}\right),
    \label{startofappend}
\end{equation}
where the covariant derivative term is given as
\begin{equation}
\begin{aligned}
    \nabla_\alpha (D_\rho {e}_{k \tau}) = &+D_\alpha D_\rho e_{k \tau}-\Gamma_{\alpha \rho}^a D_a e_{k \tau} \\
    &-\Gamma_{\alpha k}^a D_\rho e_{a \tau}-\Gamma_{\alpha \bar{\tau}}^{\bar{b}} D_\rho e_{k b},
    \label{covfirstorder}
\end{aligned}
\end{equation}
and the connections up to the first order
\begin{equation}
\begin{aligned}
    &\Gamma_{\alpha \rho}^a = \frac{1}{2} G^{ac} D_\alpha e_{\rho c}, \quad  \Gamma_{\alpha k}^a = \frac{1}{2} G^{ac} D_\alpha e_{k c}, \\
    &\Gamma_{\alpha \bar{\tau}}^{\bar{b}}= \frac{1}{2}G^{bd}(D_\alpha e_{d \tau} + \bar{D}_\tau e_{\alpha d} - \bar{D}_d e_{\alpha \tau}).
\end{aligned}
\end{equation}
Notice that when I put the terms $\Gamma_{\alpha \rho}^a$ and $\Gamma_{\alpha k}^a$ into (\ref{covfirstorder}), there will be contractions of barred and unbarred indices, so these terms are not $O(D,D)$ covariant and the first term gives the second order term. Leaving out these terms and using the definition $\Gamma_{\alpha\bar{\tau}}^{\bar{b}}$ in (\ref{startofappend}), I have 
\begin{equation}
    +\frac{1}{8} (D^{\alpha} D^{\rho} {e}_{\mu \nu})\bigg((D_\alpha e^{b \nu} + \bar{D}^\nu e_{\alpha}{}^b- \bar{D}^b e_{\alpha}{}^\nu)(D_{\rho} e^{\mu}{}_b)\bigg).
\end{equation}
When I expand this, the term with $D_\alpha e^{b \nu}$ breaks $O(D,D)$ symmetry. Then, from (\ref{startofappend}), the $O(D,D)$ invariant terms up to third order are given as
\begin{equation}
    \frac{1}{8} (D^\alpha D^\rho e_{\mu \nu })\left((\bar{D}^\nu e_\alpha{}^b)(D_\rho e^\mu{}_b)-(\bar{D}^b e_\alpha{}^\nu)(D_\rho e^\mu{}_b)\right).
    \label{appendix12}
\end{equation}
Now I move to the second case, which gives the same contribution as the first case by the symmetry of indices. So, considering first and second cases I obtain
\begin{equation}
    \frac{1}{4} (D^\alpha D^\rho e_{\mu \nu })\left((\bar{D}^\nu e_\alpha{}^b)(D_\rho e^\mu{}_b)-(\bar{D}^b e_\alpha{}^\nu)(D_\rho e^\mu{}_b)\right).
    \label{appendix1and2}
\end{equation}
The next step is to find the terms that do not come from the connections of covariant derivatives. I will start from
\begin{equation}
    -\frac{1}{4} g^{\mu k} g^{\nu \tau} g^{\alpha m} g^{\rho q}\left(\mathcal{D}_\alpha \mathcal{D}_\rho \mathcal{E}_{k \tau}\right)\left({\mathcal{D}}_m {\mathcal{D}}_q \mathcal{E}_{\mu \nu}\right)
\end{equation}
and I will put the next order terms coming from $g^{\mu k}, g^{\nu \tau}, g^{\alpha m}, g^{\rho q},\mathcal{D}_\alpha,  \mathcal{D}_\rho, {\mathcal{D}}_m,  {\mathcal{D}}_q, \mathcal{E}_{k \tau},  \mathcal{E}_{\mu \nu}$ one by one. The first observation is that the extra pieces coming from the expansions of  $g^{\mu k}, g^{\nu \tau}, g^{\alpha m}, g^{\rho q}$, $ \mathcal{E}_{k \tau},  \mathcal{E}_{\mu \nu}$ are $O(D,D)$ violating. To find the $O(D,D)$ invariant terms, I will consider the next order terms which are coming from the derivative operators using the definitions (\ref{backindepderivfinal}) up to the first order in $e_{ij}$,
\begin{equation}
 \begin{aligned}
&\mathcal{D}_i  =D_i-\frac{1}{2} e_{i k}\left(\bar{D}^k-D^k\right), \\
&\bar{\mathcal{D}}_i  =\bar{D}_i+\frac{1}{2} e_{k i}\left(\bar{D}^k-D^k\right).
\end{aligned}
\end{equation}
Note that the $e_{ik} D^k$ term in the first line and $e_{ki}\bar{D}^k$ term on the second line are $O(D,D)$ violating and I ignore them at this point. So, from the expansion of four derivative operators, the $O(D,D)$ invariant terms up to third order take the form
\begin{equation}
    \begin{aligned}
        \frac{1}{4}e_{\alpha k}\left((D^\alpha D^\rho e^{\mu \nu})(\bar{D}^k D_\rho e_{\mu \nu})-(\bar{D}^k e_{\mu \nu})(\square D^\alpha e^{\mu \nu})\right),
        \label{appendix11}
    \end{aligned}
\end{equation}
after an integration by parts.
Finally, the last third order term will come from considering the second term in the series expansion of $e^{-2d}$
\begin{equation}
    -2d \left(-\frac{1}{4} \left(D^\alpha D^\rho e^{\mu \nu}\right)\left({D}_m {D}_q e_{\mu \nu}\right) \right) = \frac{1}{2} d (D_\alpha D_\rho e^{\mu \nu})^2.
\label{100}
\end{equation}
In the end, by summing up the terms (\ref{appendix1and2}), (\ref{appendix11}), (\ref{100}), I find (\ref{mycubic1}).

\bibliographystyle{unsrt}
\bibliography{apssamp}

\end{document}